# A Community Exoplanet Imaging Data Challenge for Roman CGI and Starshade Rendezvous

**Margaret C. Turnbull[1,*], Neil Zimmerman[2], Julien H Girard[3], Sergi R. Hildebrandt[4,5], Zhexing Li[6], Ell Bogat[2], Junellie Gonzalez-Quiles[7], Christopher Stark[2], Avi Mandell[2], Tiffany Meshkat[8], Stephen R. Kane[6]**

[1]SETI Institute, Mountain View, CA 94043, USA.

[2]NASA Goddard Space Flight Center, Greenbelt, MD zip code, USA.

[3]Space Telescope Science Institute, Baltimore MD 21218, USA.

[4]Jet Propulsion Laboratory, California Institute of Technology, Pasadena, CA 91109, USA.

[5]Division of Physics, Mathematics and Astronomy, California Institute of Technology, Pasadena, CA 91125, USA.

[6]Department of Earth and Planetary Sciences, University of California, Riverside, CA 92521, USA.

[7]Department of Earth and Planetary Sciences, Johns Hopkins University, Baltimore, MD 21218, USA.

[8]IPAC, California Institute of Technology, Pasadena, CA 91109, USA.

**Abstract.** Operating in an unprecedented contrast regime ($10^{-7}$ to $10^{-9}$), the *Roman* Coronagraph Instrument (CGI) will serve as a pathfinder for key technologies needed for future Earth-finding missions like HabEx and LUVOIR. The *Roman* Exoplanet Imaging Data Challenge (*Roman* EIDC) was a community engagement effort that tasked participants with extracting exoplanets and their orbits for a 47 UMa-like target star, given: (1) 15 years of simulated precursor radial velocity (RV) data, and (2) six epochs of simulated imaging taken over the course of the *Roman* mission. Led by the Turnbull CGI Science Investigation Team, the *Roman* EIDC was preceded by four tutorial "hack-a-thon" events in Baltimore, Pasadena, New York City, and Tokyo. The *Roman* EIDC officially launched in October 2019 and ran for 8 months, offering a unique opportunity for exoplanet scientists of all experience levels to get acquainted with realistic near-future imaging data. The *Roman* EIDC simulated images include 4 epochs with CGI's Hybrid Lyot Coronagraph (HLC) plus 2 epochs with a starshade (SS) assumed to arrive as part of a Starshade Rendezvous later in the mission. In this paper, we focus on our in-house analysis of the outermost planet "d," for which the starshade's higher throughput and lower noise floor present a factor of ~4 improvement in signal-to-noise ratio over the narrow-field HLC. We find that, although the RV detection was marginal for planet d, the precursor RV data enable the mass and orbit to be constrained with only 2 epochs of starshade imaging. Including the HLC images in the analysis results in improved measurements over RV + SS alone, with the greatest gains resulting from images taken at epochs near maximum elongation. Combining the two epochs of SS imaging with the RV + HLC data resulted in a factor of ~2 better orbit and mass determinations over RV + HLC alone. In summary, the *Roman* CGI, combined with precursor RV data and later-mission SS imaging, form a powerful trifecta in detecting exoplanets and determining their masses, albedos, and system configurations. While the *Roman* CGI will to break new scientific and technological ground with direct imaging of giant exoplanets within ~5 AU of V~5 and brighter stars, a *Roman* Starshade Rendezvous mission would additionally enable the detection of planets out to ~8 AU in those systems.

*Further author information: email: turnbull.maggie@gmail.com

# 1. Introduction

## 1.1 The *Roman* Mission and Exoplanet Coronagraph

The 2.4-m *Nancy Grace Roman Space Telescope* (*Roman*; formerly the Wide Field Infrared Survey Telescope, WFIRST)[1], is a NASA mission set to launch in mid 2020s and operate at the second Sun-Earth Lagrange point (SEL2). *Roman* will carry two scientific instruments: (1) the Wide Field Instrument, providing imaging resolution comparable to that achieved by the *Hubble Space Telescope* but with 100 times the field of view, and (2) a high-contrast, small field of view Coronagraph Instrument (CGI). The *Roman* CGI is intended to serve as a pathfinder instrument for future exoplanet imaging missions like HabEx[2] by demonstrating key starlight suppression technologies, including the first active wavefront control in space.[3] CGI aims to achieve detection limits of $10^{-7}$ (minimum requirement) to $10^{-8}$ (current best estimate) at separations of ~0.15-1.5 arcseconds from bright (V<6) host stars, which would enable imaging and low-resolution spectroscopy of giant planets in reflected starlight.[4] The simulations in the *Roman* Exoplanet Imaging Data Challenge (*Roman* EIDC) employed the most recent end-to-end observing scenario models (OS6),[5] which enabled detections down to ~$10^{-9}$. As of this writing, the *Roman* mission has passed its Preliminary Design Review and entered Phase C (final design and fabrication).

## 1.2 Goals and Scope of the Data Challenge

The goals of the *Roman* Exoplanet Imaging Data Challenge[6] were:

- Familiarize the exoplanet community with the CGI data products, and with working in a new contrast regime that enables the detection of mature giant planets in reflected starlight,
- Develop, use, and improve data simulation and analysis tools, including combining RV and imaging data and using a variety of post-processing and PSF subtraction techniques,
- Explore the detection space enabled or enhanced by combining precursor RV and late-mission starshade observations with the CGI data, and
- Train and foster collaborations between future leaders of exoplanet science.

The scope of the *Roman* EIDC was to unveil an exoplanetary system hidden in realistic CGI data which includes high fidelity wavefront control residuals, a realistic EMCCD detector model, and astrophysical contamination. Participating teams were given 6 simulated imaging epochs for a nearby star (modeled after 47 UMa), calibration files corresponding to OS6, 15 years of simulated precursor radial velocity (RV) data (described in Section 2.3), and basic stellar system information (V-band magnitude, distance, mass, spectral type, and proper motion). Participants were not told the number of planets present, nor any prior information about background stars, galaxies, or exozodiacal light. From these data, participants were asked to determine the number of planets detected, and to find their orbital parameters, masses, radii, and albedos. Our three-planet 47 UMa-analog system was designed so that the planets (b, c, and d) would be in stable orbits and fall in or out of the field of view at a various detection levels in the various epochs. In only one instance was the entire planetary system (marginally) detectable in a single epoch. The imaging and RV simulations will be discussed in detail in a future publication.[7]

## 1.3 The Challenge in 4 steps

The *Roman* EIDC was rolled out in four increasingly difficult steps, to help participants work in a logical manner and avoid overwhelming the teams. Participants submitted the results of each step in order to receive the next package of simulated data:

**Step 1:** *HLC only.* Identify all the point sources in 4 HLC epochs, disentangle planets from any background sources, provide system census and astrometry.

**Step 2:** *HLC + RV priors.* Compute orbital parameters and masses using the above 4 HLC epochs plus priors from RV data.

**Step 3:** *HLC + RV + Starshade.* Refine the orbital parameters and masses of all detected planets using two additional SS epochs.

**Step 4:** For each planet, measure the phase curve (assuming a Lambertian reflectance function), and derive the radius and albedo from a provided mass-radius relationship.

Importantly, the SS images were not provided until Step 3, because these data would have undermined the "blind" analyses of the prior epochs by clearly revealing planets that are only marginally detected with the HLC. **Figures 1** and **3** show the science grade data products (post PSF subtraction) for the HLC and SS, respectively.

## 1.4 Hack-a-thons and a legacy tutorial suite

To engage the community prior to launching the *Roman* EIDC, we held four tutorial "hack-a-thon" events in Baltimore, Los Angeles, New York and Tokyo. Hack-a-thon participants were given two rehearsal data sets as well as a suite of Jupyter notebooks (mainly written in Python) for the participants to begin working with the data. This training material was developed and improved over the duration of the challenge, and is now available for public use.[8] Altogether, a diverse group of over 70 people attended the hack-a-thons, a subset of whom ultimately participated in the full data challenge either individually or in teams. In an upcoming paper,[9] we will report on the *Roman* EIDC results in detail.

## 2. Data simulations

### 2.1 HLC images

The baseline imaging mode of CGI is a Hybrid Lyot Coronagraph (HLC) operating in a bandpass centered at 575 nm (the 546 – 604 nm "Band 1" filter of CGI). This configuration provides the smallest inner working angle of the available coronagraph modes, approximately 150 mas, which results in the best overall sensitivity for detecting exoplanets in reflected starlight. The data challenge files include four epochs of simulated HLC images of the scientific target at time intervals $\Delta T$ = 0.0, 0.15, 1.0, and 2.0 years from an initial observation on 2026 Nov 1. The time-varying residual starlight pattern, including speckles and pointing jitter effects, is based on the "Observing Scenario 6" (OS6) PSF time series prepared by the CGI project's integrated modeling team in 2018.[5] The OS6 PSF time series includes alternating observatory roll angles, to assist in discriminating astrophysical signals from quasi-static speckles. Therefore, the simulated HLC images are provided as co-added images with total integration time 66000 seconds (approximately 18 hours) in each of two roll angles 26 degrees apart. Along with the science target observation. None of the provided images have been post-processed, and it was left to the participants to apply either reference differential imaging (RDI) or angular differential imaging (ADI) to partly subtract the residual starlight pattern.

The result of our "in-house" image reduction applied to two simulated HLC data epochs is shown in **Figure 1**. There are two relatively bright point sources, one of which is a planet (NE quadrant, near the occulting mask) and the other which is a background star (SW quadrant, near the edge of the field). In addition, the reduced images reveal diffuse scattered light from a debris disk with a peak surface brightness of 19.5 mag arcsec$^{-2}$ in *V*-band (equivalent to 10 "zodi" surface brightness units). Lastly, the source that we highlight for this article is a faint outer planet, circled in green in **Figure 1**. This planet "d" source would be indistinguishable from speckle noise if these HLC images were the only evidence for its existence. However, planet d reappears as a much higher signal-to-noise ratio (SNR) point source in the later starshade rendezvous images (Section 2.2), with motion consistent with a Keplerian orbit and also physically consistent with a weak, long-period residual feature in the data challenge radial velocities (Section 2.3).

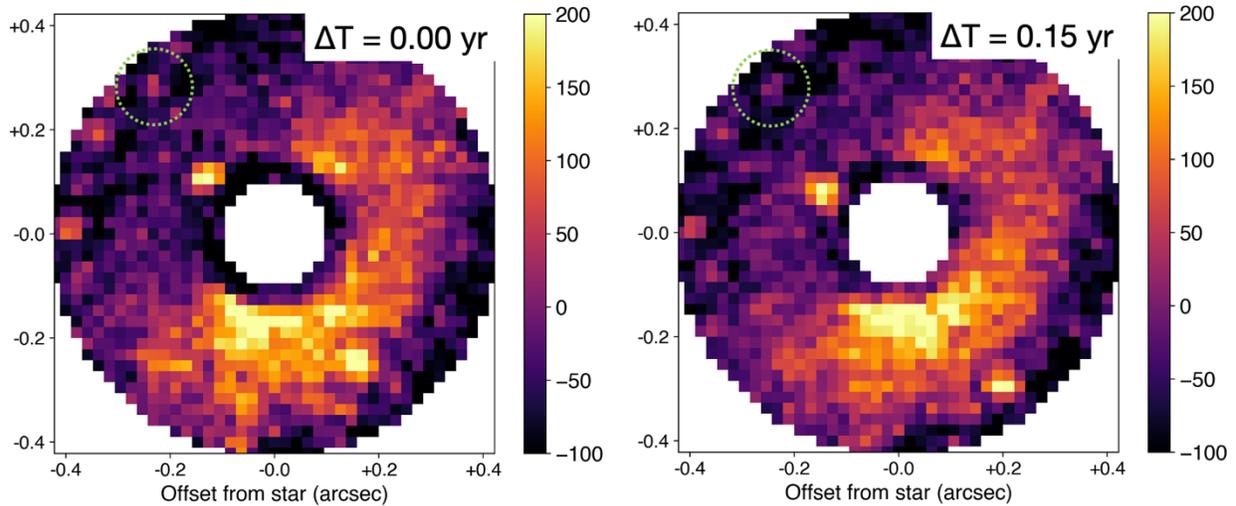

**Figure 1**: Simulated CGI HLC data after applying reference differential imaging and roll combination to the co-added images at two epochs, displayed in units of integrated photoelectrons. The faint source corresponding to planet "d" is circled in green. The data analysis presented in this paper was carried out "in house" by the Roman CGI Turnbull Science Investigation Team members.

## 2.2 Starshade images

The starshade images were generated with SISTER,[10] an open source software capable of performing starshade simulations with high optical fidelity. We chose the 425 – 552 nm band of the Starshade Rendezvous Probe (SRP)[11], see **Figure 2**, because it is the closest channel to the HLC simulated images. The simulation uses nominal instrument parameters consistent with *Roman* end-of-life (EOL) operation and includes epochs consistent with the required solar exclusion angles of the mission.

**Telescope and Detector.** The telescope pupil includes the secondary mirror central obscuration and struts, and is 2.36 m in diameter. The detector noise follows the expected *Roman* EMCCD model at its EOL. The particular values of the different parameters come from the *Roman* Space Telescope parameters listed online.[12] The readout noise is 100 e- per frame, EM gain is 1,000, dark current at EOL is 0.77 e-/pix/hour, and clock induced charge noise is 0.02 e$^-$/pix/frame. In order to be able to compare the contribution of starshade data to the CGI data more easily, the total integration time set for both epochs with starshade data was 1.5 days, the same as for the CGI HLC epochs. The detector Quantum Efficiency is based on laboratory measurements of *Roman*'s EMCCD detector. The effective QE across the 425-552 nm is 0.45 and it is the result of the actual QE and further losses due to cosmic rays, charge transfer efficiency, and hot pixels. The pixel scale in the *Roman* coronagraph instrument is 21.85 mas/pixel. We have assumed that all the pixels are identical without any hot pixels. The simulation does not include contamination from cosmic rays or sources of noise not intrinsic to the detector electronics. The end-to-end optical throughput is the product of reflection losses in the telescope (0.81), the coronagraph instrument (0.60, exclusive of coronagraph masks which are not used), and the starshade dichroic filter (0.9) for a net throughput of 0.44.

**The Starshade.** The starshade has 24 petals and is 26 m in diameter. The starshade's geometric inner working angle (IWA) is the apparent size of the starshade's radius as seen by the *Roman* telescope and in the 425-552 nm passband is 72 mas, as shown in **Figure 2**. For sources located far from the starshade's IWA, the optical response (PSF FWHM = 40.6 mas at mid-band) is that of the *Roman* telescope, including the secondary mirror and 6 supporting struts.

**Starshade Imperfections.** In this paper we define "contrast" as the ratio of the intensity averaged over an image plane resolution element (typically the PSF FWHM) to the peak intensity of the target star when the starshade is not present. The current best estimate of the starshade contrast in the SRP report[11] is 4x10$^{-11}$. In the *Roman* EIDC, we considered an imperfect starshade that would allow us to simulate possible manufacturing and deployment errors of the starshade. The SRP report sets the error budget of the starshade contrast to $1.0 \times 10^{-10}$. However, the locations of the planets on the image plane fall well outside the 72 mas IWA for all of them (the closest angular separation being 116 mas). At those angular distances, the contrast of the starshade improves substantially. In order to explore any potential effect of the residual starlight on the planets' analysis, we artificially degraded the performance of the starshade to an average contrast over the passband of $1.4 \times 10^{-10}$ at the IWA. This degraded contrast performance was achieved by displacing the petals in-plane as 24 independent rigid bodies with radial and azimuthal offsets such that the root-mean-square displacements of all points along the petals was 0.56 mm. We did not include shear displacement of the starshade relative to the line of sight between the telescope and target star (expected to be <2 mas), nor any tilt of the starshade plane with respect to the telescope. Both of these effects have negligible impact on the exoplanet analyses.[10]

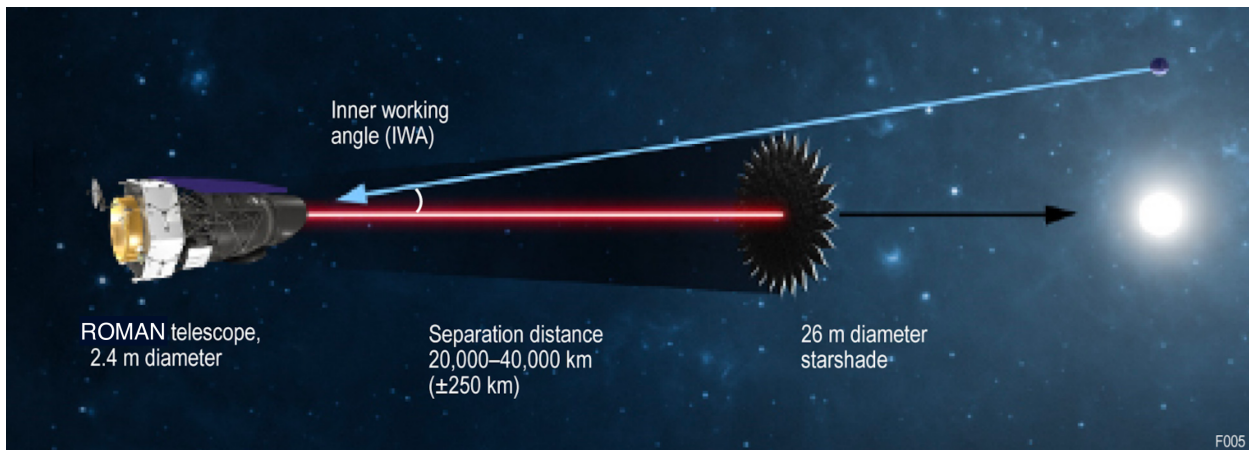

**Figure 2:** The geometry of the starshade and *Roman* telescope in the SRP mission concept.[11] The starshade's inner working angle (IWA) is the apparent size of the starshade's radius as seen by the *Roman* telescope. In the 425-552 nm passband, the IWA is 72 mas.

**Solar Glint.** For both SS imaging epochs, the sun is located 60° from the starshade normal, which is a median value among planned observations, and is oriented at 0° from horizontal. Solar glint (sunlight reflected off of the starshade into the telescope) is computed under a "worst case scenario" assumption that the starshade edges are not coated with any anti-reflection coating. The resulting solar glint intensity is therefore ~10 times greater than current best estimates, in order to better explore its impact on the data analysis.

**Astrophysical Components.** The astrophysical components of the SS simulations include the three-planet system itself, plus residual starlight from an imperfect starshade, solar glint due to sun's light scattered through the starshade petals, local zodiacal light, exozodiacal dust light, and a background galaxy. The planets' apparent separations for epoch $\Delta T = 3$ years are 116.6, 271.0 and 488.2 mas, respectively, and for epoch $\Delta T = 4$ years they are 159.6, 182.7 and 502.1 mas. In all cases, the projected distance is significantly larger than the geometric IWA of the starshade in this configuration (72 mas).

We assumed a Lambertian phase function to derive the intensity of the reflected light from the planets, which is good enough to fulfill the aims of the data challenge. The flux ratios for epoch $\Delta T = 3$ years are $5 \times 10^{-10}$, $5.6 \times 10^{-10}$ and $6.7 \times 10^{-10}$, for planets b, c, and d, respectively, and for epoch $\Delta T = 4$ years they are $2.4 \times 10^{-9}$, $1.2 \times 10^{-10}$ and $4.7 \times 10^{-10}$, respectively. For comparison, Earth's flux ratio at quadrature is $1.1 \times 10^{-10}$ at 500 nm. The planet orbital configuration was chosen to provide low flux ratios during the starshade epochs to better demonstrate the capability of SS imaging.

The exozodiacal cloud is assumed to have a density 5 times that of the solar system's zodiacal cloud. Its dust distribution was derived from an N-body dynamical simulation of dust particles under the gravitational influence of a co-orbiting planet.[13] The dust grains in this model tend to become trapped in mean motion resonances with a planet. In the data challenge images, the scattered light from these resonant dust structures produces diffuse intensity enhancements that co-orbit with the innermost planet "b." For the extragalactic background object, we used the Haystacks data cubes[14] and selected the brightest background galaxy from the HST deep field survey. In Section 3, we provide the relative contribution of each component for the case of planet d.

Simulated images for the two epochs with the starshade without any post-processing are shown in **Figure 3**. Both images show planet d clearly inside the dotted circle. In epoch $\Delta T = 3$ year, planet c is located near (-0.10, -0.25) arcsec, and can be seen as a faint perturbation to the exozodiacal dust, while planet b is masked by a bright clump of the exozodiacal light. In epoch $\Delta T = 4$ year, planet b is clearly visible near (0.10, 0.15) arcsec, while planet c is masked by the exozodiacal light. The images also show how the resonant structure of the exozodiacal revolves around 47 UMa from one epoch to the following one, and its impact on planet detection.

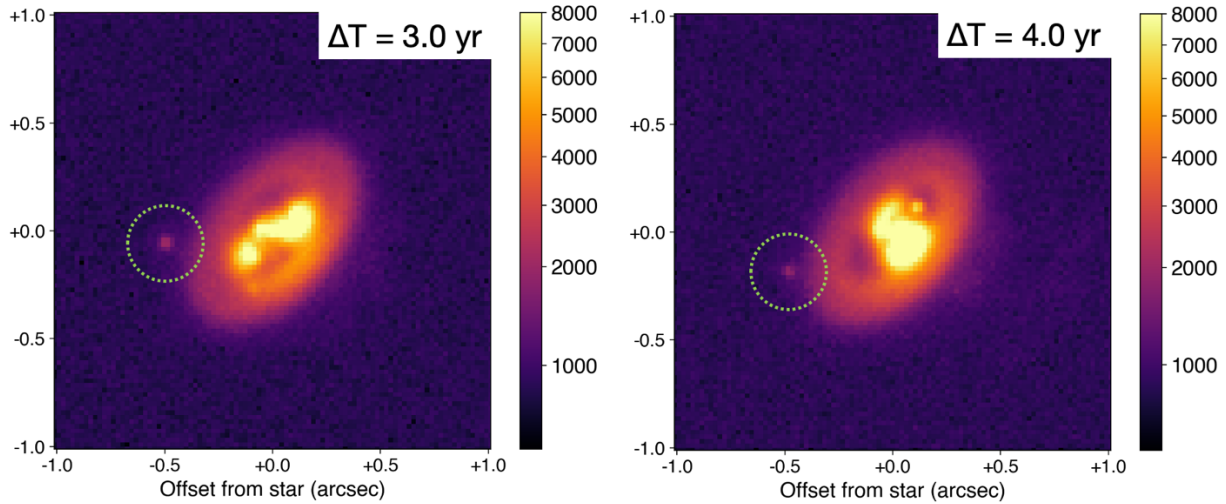

**Figure 3**: Co-added *Roman* CGI/Starshade images of the data challenge target, without any post-processing, displayed in units of integrated photoelectrons. The point source corresponding to planet "d" is circled in green.

## 2.3 Precursor Radial Velocity data

Synthetic RV data points were created spanning 15 years, first 10 years with RV instrumental precision of 1 m/s similar to instruments such as Keck HIRES, and additional 5 years with 0.3 m/s taking into account the emerging extreme precision RV instruments such as the WIYN NEID spectrograph. A total of 200 data points were created by solving Kepler's equation and superposing individual signals from each planet. The number of data points and the length of coverage is similar to the archival RV coverage for 47 UMa, except that the data points are spread out nearly evenly (i.e., in the most optimal way). No planet-planet interactions are included in the RV data, since such effects are not observable with RV.

To mimic different sources of noise, each RV data point was passed through a Gaussian filter to randomize the velocity measurements, with the sigma calculated from the quadrature sum of instrumental error and stellar jitter. The jitter value of around 2.1 m/s was estimated using the method from Isaacson & Fischer[15] assuming the stellar mass and S-index values of 47 UMa host star. The total RV as well as individual phase-folded RV for each planet is shown below. The figure is generated using RV modeling toolkit *RadVel* [16] with all the parameters fixed so that the outermost planet can be properly displayed. Parameters for each planet such as the orbital period (P), semi-amplitude (K), and eccentricity (e) are shown in the upper right corner of each phase-folded panel.

## 3. Revealing planet d with the starshade

We designed the fictitious data challenge planetary system to probe a range of detection difficulties. The outermost planet "d" was chosen to represent a particularly stressing case for CGI's sensitivity limits: a roughly Saturn-sized planet at a 7.7 AU semi-major axis, with a visible wavelength geometric albedo of 0.5. At a resulting planet-to-star flux ratio of approximately $10^{-9}$, the intensity of planet d is almost equal to the residual speckle noise standard deviation in the HLC image after classical RDI and roll combination. Therefore, in the context of HLC data alone, planet d does not present an obvious source candidate among the various residual subtraction features. However, when analyzed in concert with the starshade images and supporting radial velocities, a more compelling picture emerges.

Owing to the higher throughput and much lower systematic noise of the Roman/Starshade observing configuration, the SNR of the planet d point source increases from SNR~1 to SNR~4 in the starshade imaging epochs ($\Delta T = 3$ years and $\Delta T = 4$ years). Due to its projected location for outside the IWA on the image plane (488.2 mas in $\Delta T = 3$ years and 502.1 mas in $\Delta T = 4$ years compared to the 72-mas IWA), the contributions from the residual starlight and solar glint to planet d's signal are negligible compared to other background sources. Within a circular area with a diameter equivalent to 1 FWHM, the background source contribution is approximately 30% of planet d's signal. Furthermore, 94% of the background source is due to local zodiacal light and 6% to the exozodiacal cloud light. The spatial distribution of the local zodiacal light is very smooth at the angular scales of these images, and it can easily be subtracted, leaving only the shot noise from the exozodi signal. As mentioned before, the overall SNR of planet d is ~4.

These two starshade point source detections (one in each starshade epoch), considered in combination with the physically consistent long-period residual sinusoid in the radial velocities (**Figure 4**), would strongly motivate a retrospective search for a planet d signal in the earlier HLC data. Scenarios like this one -- where a new detection is reinforced by the recovery of a previously missed signal in older data -- are common in observational astronomy. One famous example in high-contrast imaging is the detection of the HR 8799 planets b, c, and d in archival Hubble Space Telescope data, in NICMOS images acquired 10 years before the discovery of those objects.[17]

The regions containing the planet d signal in the first two HLC images are circled in **Figure 1**. To carry out our "in-house" astrometry analysis, we measured the point source centroids by cross-correlating a PSF model with a small cutout centered on the signal region. Our astrometry data points are plotted as the filled-in circles in **Figure 5**, alongside different symbols representing the astrometry reported by data challenge participants, whose results will be described in a forthcoming publication. Note that our in-house analysis could not recover even a marginal signal in the 3rd epoch HLC image.

Furthermore, since planet d passes outside of the HLC outer working angle after this time, no detection is possible in the 4th imaging epoch ($\Delta T = 2$ years).

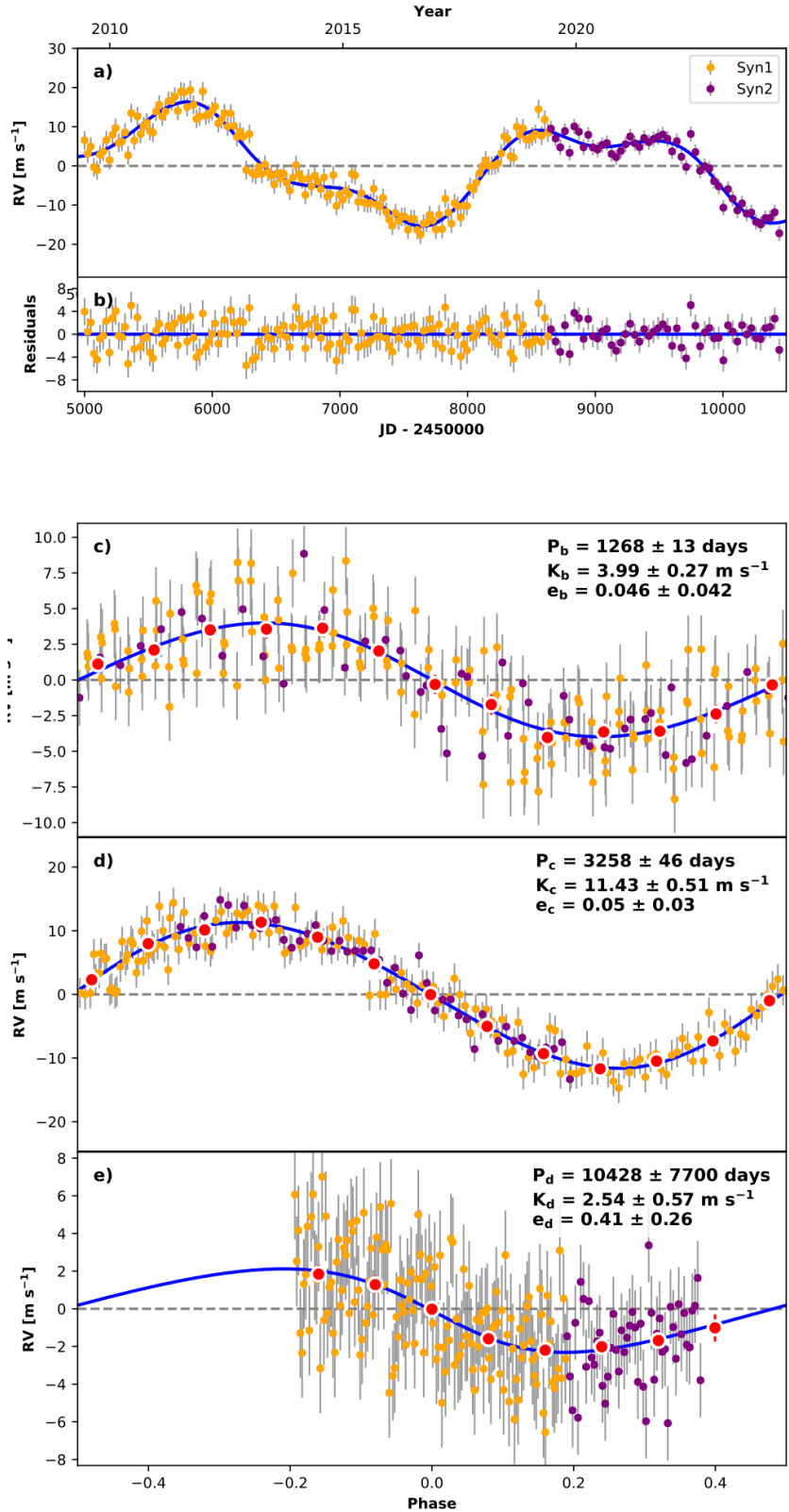

**Figure 4**: Synthetic RV for the system. Upper panel (a) and (b) show the total RV with the best fit 3-planet model along with residuals. Lower panels (c), (d), and (e) are individual phase-folded RVs for each planet.

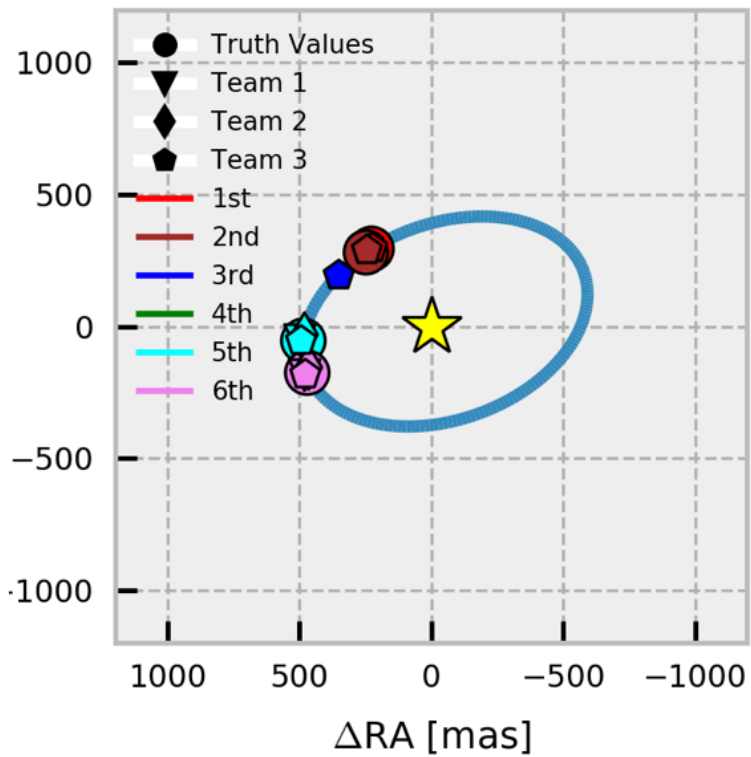

**Figure 5**: Astrometry of planet d reported by data challenge participants. The true projected Keplerian orbit is shown by the blue ellipse. The first two epochs are very low (SNR ~ 1) HLC detections that would most likely only be recovered based on retrospective analysis after eventual *Roman* Starshade Rendezvous detections.

## 3.1 Orbit fitting

By combining the two epochs of starshade astrometry (**Figure 3**) with the residual sinusoidal feature in the radial velocities (third panel of **Figure 4**), it is possible to place some preliminary constraints on the Keplerian orbit. In our orbit fitting trials with the open source *RadVel* [16] and *orbitize!* [18] packages, we found that when applying priors on period, eccentricity, and ascending node position angle, with the two epochs of starshade astrometry, *orbitize!* constrains the semi-major axis to a 68% confidence interval of 7.4--10.5 AU, and the inclination angle to a 68% confidence interval of 41–56 deg.

With the addition of the weaker point source detections in the HLC images, the fit is improved by the increased time baseline. In **Table 1**, we list the 68% confidence intervals of the inclination angle and planet mass retrieved from the combined orbit fit, alongside the true values used to generate the simulated data. The planet mass constraints are modest, being limited by the quality of the RV signal and the fact that the imaging astrometry only covers ⅕ of planet d's 21-year orbital period. Even with all 4 epochs of astrometry, the confidence interval spans more than a factor of 2 in mass, from 0.24 Jupiter masses ($M_{Jup}$) to 0.55 $M_{Jup}$. The other two data challenge planets are situated at smaller orbital radii, with semi-major axes 2.3 AU and 4.2 AU. In those cases, the narrower confidence intervals of the RV semi-amplitudes, along with larger orbital period coverage of the imaging data, reduce the final mass uncertainties to within 20% of their true values. We will describe the results for these other data challenge planets, with emphasis on the participants' analysis, in forthcoming publications. Because those inner planets are brighter and have shorter orbital periods, the HLS data are sufficient to detect and estimate their complete orbital parameters to similar precision, without starshade observations.

Table 1: Estimated orbital inclination and mass of planet d, based on combining radial velocity priors with various subsets of astrometry. These measurements are from an "in-house" analysis of the simulated data; results from the data challenge participants are deferred to forthcoming publications.

|  | 2 epochs (HLC only) | 2 epochs (Starshade only) | 4 epochs (HLC and starshade) | True value |
|---|---|---|---|---|
| inclination (deg) | $71^{+38}_{-28}$ | $50^{+6}_{-9}$ | $48^{+5}_{-6}$ | 48 |
| mass ($M_{Jup}$) | $0.25^{+0.33}_{-0.10}$ | $0.34^{+0.24}_{-0.12}$ | $0.35^{+0.20}_{-0.11}$ | 0.32 |

## 3.2 Photometry and phase curve analysis

One of the novel aspects of future reflected starlight direct imaging will be the ability to observe the photometric variation over the course of the orbit due to the changing star-planet-observer phase angle. Combined knowledge of the orbit, along with an inferred planet radius, will enable observers to estimate the bulk geometric albedo from a flux ratio time series. As a demonstration, we used our imaging photometry of planet d, in combination with the previous orbit fit, to estimate the planet's geometric albedo, assuming a Lambertian sphere scatterer.

Before fitting for the planet's albedo, we first used the posterior mass estimate from the orbit fit and radial velocity semi-amplitude, and then inferred the planet's radius from an empirical mass-radius relationship along with the approximate equilibrium temperature of planet d.[19] A simple least-squares optimization finds the albedo that minimizes the distance between the 4 epochs of photometry and the Lambertian phase curve associated with the orbit posteriors. The result, in **Figure 6**, illustrates the four photometric data points, the best-fit flux curve, and the flux curve of the true planet parameters.

One of the dramatic features shown in Figure 6 is the improved photometric constraints enabled by the higher SNR starshade detections. A slight bias in the starshade photometry causes the albedo to be underestimated (0.3 versus the true value of 0.5), but this bias is still much smaller than the overall uncertainty of the albedo estimate. The fit shown in Figure 6 corresponds only to the best estimated orbit parameters and mass, not the envelope of all photometric phase curves within the confidence intervals of all orbital elements. If we repeat the fit at the minimum and maximum extremes of the 68% confidence intervals of all the orbit parameters derived from the data, then the albedo confidence interval spans 0.04 to 0.73. A longer time series with photometry at the starshade precision would enable better constraints.

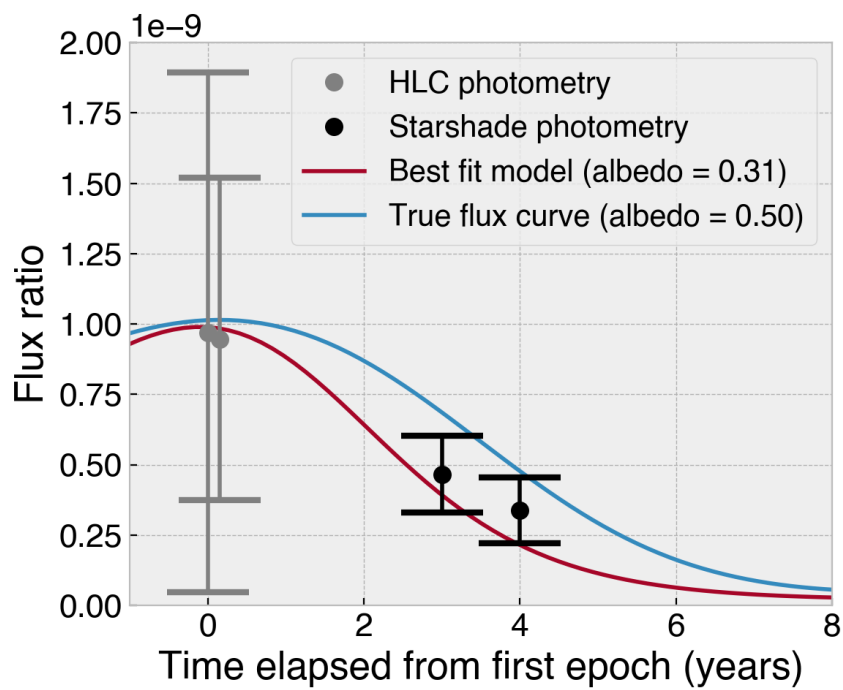

**Figure 6**: Best fit (red) to the geometric albedo of planet d using photometry and orbit posteriors, assuming a Lambertian sphere scatterer. The true flux curve is shown in blue.

## 4. Summary

The *Roman* CGI, combined with precursor RV data and late-mission SS imaging, form a powerful trifecta in detecting exoplanets and determining their masses, albedos, and system configurations. While the CGI is more nimble than a starshade and can more readily target specific epochs to reduce uncertainty in planet orbits, the higher throughput of the starshade enables detection and study of planets much farther out in the system and resonant exozodiacal dust structures. In the *Roman* CGI Exoplanet Data Challenge, we found that even a marginal prior RV detection of outermost planet d enabled its mass and orbit to be constrained with only 2 epochs of starshade imaging. Including the HLC images in the analysis results in improved measurements over RV + SS alone, with the greatest gains resulting from images taken at epochs near maximum elongation. Combining the two epochs of SS imaging with the RV + HLC data resulted in a factor of ~2 better orbit and mass determinations over RV + HLC alone. While the *Roman* CGI is expected to break new ground with direct detection of giant exoplanets within ~5 AU of V~5 and brighter stars, a *Roman* Starshade Rendezvous mission would additionally enable the detection and characterization of planets out to ~8 AU in those systems.

The *Roman* CGI Exoplanet Data Challenge has proven to be an excellent way to (1) engage the community in working with the intricacies of the first mission to perform wavefront control in space, and (2) motivate the community to invest in larger missions that will enable the study of mature exoplanet systems, including habitable planets, in reflected starlight. This effort also generated many interactions between open source package developers (e.g. *orbitize!*, RadVel) and a diverse group of exoplanet scientists running them, resulting in (1) key enhancements to those packages, and (2) the creation of a legacy tutorial suite that is now available online for future training exercises.

## 5. Acknowledgments

This work was funded by NASA Grant NNG16PJ27C, which supports the Turnbull Roman CGI Science Investigation Team. We thank the Hayden Planetarium and Jackie Faherty for an excellent tour of known exoplanet host stars, and the Flatiron Institute for hosting the New York City hack-a-thon event.  We thank STScI, IPAC, and Motohide Tamura and Masayuki Kuzuhara (University of Tokyo & Astrobiology Center)  for hosting hack-a-thons in Baltimore, Pasadena, and Tokyo.  We thank John Krist and the JPL project science team for the OS6 simulations, and we thank Sarah Blunt for her many interactions with the *Roman* EIDC team and participants and for her ongoing work enhancing the extremely useful *orbitize!* package.  We also thank BJ Fulton for contributing his expertise with RadVel.  Finally, we thank all of the Hack-a-thon and Data Challenge participants for joining us in this community venture and providing extensive feedback over the last two years.